# Comparison of some commonly used algorithms for sparse signal reconstruction

Gojko Ratković, Milan Rešetar, Svetlana Zečević
Faculty of Electrical Engineering
University of Montenegro Podgorica, Montenegro

Emails: gojko.ra77@gmail.com, peso.rakov@gmail.com,
zecevicsvetlana99@gmail.com

*Abstract*—**Due to excessive need for faster propagations of signals and necessity to reduce number of measurements and rapidly increase efficiency, new sensing theories have been proposed. Conventional sampling approaches that follow Shannon-Nyquist theorem require the sampling rate to be at least twice the maximum frequency of the signal. This has triggered scientists to examine the possibilities of creating a new path for recovering signals using much less samples and therefore speeding up the process and satisfying the need for faster realization. As a result the compressive sensing approach has emerged. This breakthrough makes signal processing and reconstruction much easier, not to mention that is has a vast variety of applications. In this paper some of the commonly used algorithms for sparse signal recovery are compared. The reconstruction accuracy, mean squared error and the execution time are compared.**

*Keywords- compressive sensing; signal recovery; OLS, OMP, gradient-based algorithm*

## I. Introduction

Compressive Sensing (CS) [1]-[3] is relatively new sampling theory, opining that it could be possible to overshadow the traditional limitations of sampling theories based on Shannon-Nyquist approach. It is well known that reconstruction based on this theory requires the signal to be uniformly sampled at a frequency that is equal or higher twice the maximal signal frequency. This requirement is demanding when dealing with high frequency signals.

Those are some reasons that encouraged researchers to contemplate about this significant issue and come up with the solution that is continuously bringing progress in a number of real applications.

Compressive sensing allows signal reconstruction based on a limited number of measurements, using complex mathematical algorithms [1]-[24]. Those algorithms can roughly be divided in three main categories: Convex optimization, Greedy algorithms and Hard thresholding algorithms [3]-[5], [9], [12], [15], [16].

Furthermore, the arrangement of this paper is described in following order: Section II provides some basic information about CS, in section III aforementioned algorithms are described with proper mathematical background. Results of comparison are given in the section IV and finally short conclusion is provided at the end.

## II. Fundamentals of the Compressive Sensing

The compressive sensing theory claims that the signal can be recovered with the usage of significantly low set of randomly acquired samples alluding to the fact that the signal has so called sparse representation in a certain transform domain. Also the acquisition procedure should satisfy the random sampling. Sparsity refers to the requirement that the signal is sparse, i.e if it has only a few number of coefficients that are sufficiently different from zero in a transform domain.

Let signal $\mathbf{m}$ be an *N*-dimensional signal, which satisfies the sparsity in transform domain, $\Psi \in R^{N \times N}$. The sparse representation of signal $\mathbf{m}$ over the basis $\Psi$ is presented with the vector $\mathbf{u}$. Now, $\mathbf{m}$ can be described as

$$\mathbf{m} = \Psi \mathbf{u} \qquad (2)$$

Let $\Psi$ be the inverse Fourier transform matrix, therefore u can be interpreted as frequency domain representation of the time domain signal $\mathbf{m}$. Signal $\mathbf{m}$ could be considered to be *K*-sparse in the $\Psi$ domain if there are only $K(K \ll N)$ out of $N$ coefficients in $\mathbf{u}$ that significantly differ from zero. It is worth mentioning that sparsity is one of the main conditions that needs to be fulfilled in order to approach to signal reconstruction. The set of completely random measurements are taken from signal $\mathbf{m}(N \times 1)$, that can be represented by using random measurement matrix $\Phi(M \times N)$ as shown in eq. (3):

$$\mathbf{l} = \Phi \mathbf{m} \qquad (3)$$

Another valuable requirement that matrices $\Psi$ and $\Phi$ have to follow in order to make the compressive sensing application possible is the incoherence. The correrspondence between the number of non-zero samples in the transform domain $\Psi$ and number of needed measurments relies on the coherence between those two matrices. Having satisfied those two conditions, signal is prepared to approach to recovery using different algorithms which will be observed in text down below.

## III. Recovery Algorithms

In this work we put emphasis on Greedy and Convex Optimization algorithms, specifically on Gradient Pursuit Algorithm [9], Nearly Orthogonal Matching Pursuit [23], [24], Orthogonal Least Squares algorithm, Adaptive Gradient based Algorithm [16], [17] and L1-Minimization Algorithm With Equality Constraints. Additionally, we have implemented the Hard thresholding algorithm that keeps exactly *M* elements in each iteration (*M* will be explained in further text). We observed the sinusoidal multicomponent signal, that can be mathematically described as follows:

$$x(n) = \sum_{k=1}^{K} a_k e^{-j2\pi kn}, \quad n \in [0, N-1], \quad (1)$$

where *K* represents number of sinusoidal components in the mentioned signal, and *N* is length of the signal. The reconstruction of a signal that satisfies Eq. (1) is provided using those six algorithms.

The algorithms used in this work approximate a vector **m** iteratively. The process of iteration implies that in iteration *n* an approximation is calculated using the following formula:

$$\hat{\mathbf{m}}^n = \mathbf{\Psi}_{\Gamma^n \Pi^n}, \quad (4)$$

afterwards the approximation error is calculated as follows:

$$\mathbf{r}^n = \mathbf{m} - \hat{\mathbf{m}}^n. \quad (5)$$

In every iteration this error is used to pick a new element that will be chosen from $\mathbf{\Phi}$ in order to find the best approximation.

## A. (Nearly) Orthogonal Matching Pursuit

Orthogonal Matching Pursuit [3], [4], [23] evolved from Matching Pursuit as an improvement. Hence, they share many of the properties. So OMP finds the optimum signal approximation attainable with chosen atoms. Its algorithm is given

1) *Initialise* $\mathbf{r}^0 = \mathbf{m}, \mathbf{l}^0 = 0, \Gamma^0 = \varnothing$
2) *for* $n = 1; n \coloneqq n+1$ *till stopping criterion is met*
    *a)* $\mathbf{g}^n = \mathbf{\Phi}^T \mathbf{r}^{n-1}$
    *b)* $i^n = \arg_i \max |\mathbf{g}_i^n|$
    *c)* $\Gamma^n = \Gamma^{n-1} \cup i^n$
    *d)* $\mathbf{l}^n = \mathbf{\Phi}_{\Gamma^n}^{\dagger} \mathbf{m}$
    *e)* $\mathbf{r}^n = \mathbf{m} - \mathbf{\Phi} \mathbf{l}^n$
3) *Output* $\mathbf{r}^n, \mathbf{l}^n$

where, the † indicates the Moore-Penrose pseudo-inverse.

## B. Orthogonal Least Squares

The selection step implemented in OLS [23], [24] diverges from the one used in OMP because it selects the vector $\mathbf{\Phi}$ that results in accomplishing the minimum residual error. OLS algorithm is presented in the following order:

1) *Initialise* $\mathbf{r}^0 = \mathbf{m}, \mathbf{l}^0 = 0, \Gamma^0 = \varnothing$
2) *for* $n = 1; n \coloneqq n+1$ *till stopping criterion is met*
    *a)* $\mathbf{g}^n = \mathbf{\Phi}^T \mathbf{r}^{n-1}$
    *b)* $i^n = \arg_i \min_{\Gamma^n : \Gamma^n = \Gamma^{n-1} \cup i} \|\mathbf{m} - \mathbf{\Phi}_{\Gamma_i^n} \mathbf{\Phi}_{\Gamma_i^n}^{\dagger} \mathbf{m}\|_2$
    *c)* $\Gamma^n = \Gamma^{n-1} \cup i^n$
    *d)* $\mathbf{l}^n = \mathbf{\Phi}_{\Gamma^n}^{\dagger} \mathbf{m}$
    *e)* $\mathbf{r}^n = \mathbf{m} - \mathbf{\Phi} \mathbf{l}^n$
3) *Output* $\mathbf{r}^n, \mathbf{l}^n$

## C. Gradient Pursuit

In iteration *n* the solution made by OMP is the minimization over $\mathbf{y}_{\Gamma^n}$ of the quadratic cost-function

$$\|\mathbf{m} - \mathbf{\Phi}_{\Gamma^n \Pi^n}\|_2^2 \quad (6)$$

Instead of that in the Directional Pursuit Framework family, where GP belongs, directional update is done via:

$$\mathbf{l}_{\Gamma^n}^n = \mathbf{l}_{\Gamma^n}^{n-1} + \mathbf{a}^n \mathbf{d}_{\Gamma^n}, \quad (7)$$

where $d_{\Gamma^n}$ represents an update direction. Various directions $d_{\Gamma^n}$ can be selected and we will discuss one that belongs to GP. Furthermore, the step-size $a^n$ is defined by

$$\mathbf{a}^n = \frac{\langle \mathbf{r}^n, \mathbf{c}^n \rangle}{\|\mathbf{c}^n\|_2^2}, \quad (8)$$

Where $c^n$ is the vector $c^n = \mathbf{\Phi}_{\Gamma^n} d^n$. The Directional Pursuit Family of algorithms can be mathematically explained :

1) *Initialise* $\mathbf{r}^0 = \mathbf{m}, \mathbf{l}^0 = 0, \Gamma^0 = \varnothing$
2) *for* $n = 1; n \coloneqq n+1$ *till stopping criterion is met*
    *a)* $\mathbf{g}^n = \mathbf{\Phi}^T \mathbf{r}^{n-1}$
    *b)* $i^n = \arg_i \max |\mathbf{g}_i^n|$
    *c)* $\Gamma^n = \Gamma^{n-1} \cup i^n$
    *d) calculate update direction* $\mathbf{d}_{\Gamma^n}$
    *e)* $\mathbf{c}^n = \mathbf{\Phi}_{\Gamma^n} \mathbf{d}_{\Gamma^n}$
    *f)* $\mathbf{a}^n = \frac{\langle \mathbf{r}^n, \mathbf{c}^n \rangle}{\|\mathbf{c}^n\|_2^2}$
    *g)* $\mathbf{l}_{\Gamma^n}^n = \mathbf{l}_{\Gamma^n}^{n-1} + \mathbf{a}^n \mathbf{d}_{\Gamma^n}$
    *h)* $\mathbf{r}^n = \mathbf{r}^{n-1} - \mathbf{a}^n \mathbf{c}^n$
3) *Output* $\mathbf{r}^n, \mathbf{l}^n$

For GP using the following gradient as the update provides the minimum of (6) in a single step which is optimum for best reconstruction result.

$$g_{\Gamma^n} = \mathbf{\Phi}_{\Gamma^n}^T (\mathbf{m} - \mathbf{\Phi}_{\Gamma^n} \mathbf{l}_{\Gamma^n}^{n-1}). \quad (9)$$

## D. Adaptive Gradient Based Reconstruction Method

Adaptive Gradient Based Algorithm [14], [16], [17] belongs to the family of convex optimization approaches. After choosing initial values of the available signal samples it proceeds by iteratively varying values of aforementioned samples resulting in improved concetration in the sparsity domain. The process consists of changing the initial value for some adaptable step $\pm\Delta$ gradually approaching to the exact value. The missing signal samples can be considered as variables with zero initial value, eventually they are being updated by the gradient vector which is extracted from a difference between $l_1$ - norms of the vectors, with the slight difference of being changed for $+\Delta$ and $-\Delta$. Obtained gradient value is used to update the values of missing samples. The measure of concentration in the transform domain is used to calculate its value, as showed in following formula:

$$\eta(\mathbf{m}) = \sum_{k=0}^{N-1} |\mathbf{m}(k)| \quad (10)$$

### E. L1-Minimization Algorithm With Equality Constraints

Another member of convex optimization approaches is $l1$ with equality constrains. By using nonlinear recovery algorithms super-resolved signals and even images can be recovered from practically incomplete data. This algorithm is based on a fact that the program $(P_1)$ min $\|\mathbf{m}\|_1$ subject to $\mathbf{Am}=\mathbf{l}$, also known as basis pursuit, finds the vector with the smallest $l_1$-norm form

$$\|\mathbf{l}\|_1 := \sum_i |\mathbf{m}_i| \quad (11)$$

that elucidates the observations $\mathbf{l}$. If a sufficiently sparse $m_0$ exists, that satisfies the $\mathbf{A m}_0 = \mathbf{l}$, then $(P_1)$ will be able to find it. Furthermore, when $\mathbf{m,A}$ and $\mathbf{l}$ have real-valued values $(P_1)$ can be recast as an LP (linear program).

### F. (Iterative) Hard Thresholding Algorithm

This partcular algorithm solves this optimization problem:

$$\min_\mathbf{l} \|\mathbf{u} - \Phi \mathbf{l}\|_2^2 + \lambda \|\mathbf{l}\|_0, \quad (12)$$

by deriving the succeeding iterative algorithm.

$$\mathbf{l}^{n+1} = V_{\lambda^{0.5}}(\mathbf{l}^n + \Phi^V(\mathbf{u} - \Phi \mathbf{l}^n)), \quad (13)$$

Where $V_{\lambda^{0.5}}$ represents the element wise hard thresholding operation:

$$V_{\lambda^{0.5}}(\mathbf{l}_i) \begin{cases} 0, & if \ |\mathbf{l}_i| \leq \lambda^{0.5} \\ \mathbf{l}_i, & if \ |\mathbf{l}_i| > \lambda^{0.5}. \end{cases} \quad (14)$$

## IV. EXPERIMENTAL RESULTS

In this section, six previously described algorithms are tested on the sparse band-limited signal consisted of 6 components. In order to solve optimization problems, the CVX, l1-magic and Sparsify Matlab toolbox have been used. The aforementioned signal is:

$$x = A_1 e^{(j2\pi 28\frac{n}{N})} + A_2 e^{(j2\pi 26\frac{n}{N})} + A_3 e^{(j2\pi 6\frac{n}{N})} + A_4 e^{(j2\pi 42\frac{n}{N})} + A_5 e^{(j2\pi 90\frac{n}{N})}$$

Where component magnitudes are:
$A_1=3.5; A_2=1.5; A_3=4.4; A_4=1.8; A_5=3; N=512$ represents the length of the signal and $n \in (1, N)$.

Noisless case is considered. Signal is reconstructed using only $M=30$ samples. Fig. 1 depicts signal $\mathbf{x}$ in Fourier Domain and fig. 2 down below represent Fourier Domain of reconstructed signal using algorithms in respective order, OMP, OLS, GP, Gradient, l1-eq and IHT.

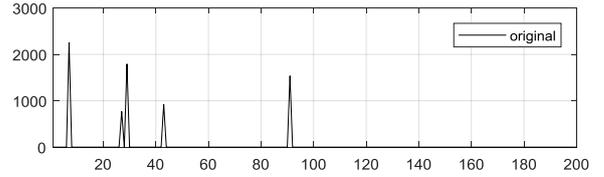

Figure 1: Fourier Domain of the original signal

Figure 1 depicts Fourier Domain of signal x which consists of five components in (DFT) domain hence making it sparse and available for reconstruction. Figure 2 shows reconstructed signals using differerent algorithms. It is obvious that the first three algorithms which belong to Greedy family are far better at reconstructing the signal than the last one that belongs to Thresholding family and l1-eq.

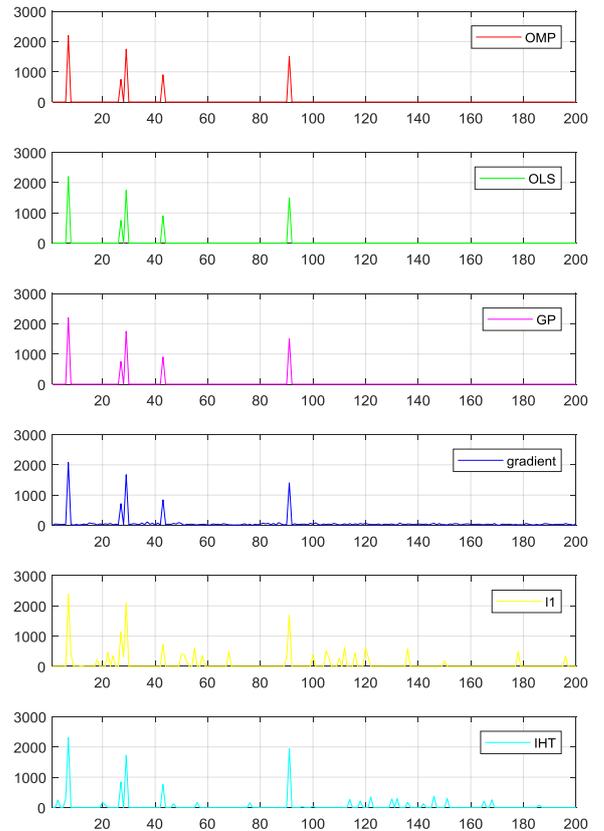

Figure 2: Fourier Domain of reconstructed signals using different algorithms

The worst performance was achieved by l1-minimization (from L1 magic software)algorithm, which produced some unwanted terms in the DFT domain. As previously mentioned that convex optimization algorithms performed better, followed by greedy algorithms, figure 3 proves this statement where we can see insufficient matching using IHT and l1-minimization. The rest of the algorithms managed to reproduce signal almost perfectly. L1-eq algorithm was able to reconstruct the exact shape of original signal, but with certain latency in time domain. As a consequence of this particular problem, l1-eq produced the biggest MSE which will be described in further text. Another comparison parameter used in this paper is calculation of MSE therefore in the figure 4 those parameters are given. Noisless signals are observed. All greedy algorithms have approximately the same MSE, which converges to 0.0127 after exceeding *M*=60 available samples. On the other hand, IHT showed a bit worse performance, where MSE converges to approximately 0.6 after reaching the exact number of available samples. Just like in other parameters, l1-eq showed the worst performance, achieving the biggest MSE. It was caused due to the fact that the signal was shifted in time compared to the original, which made this error significantly bigger compared to others. Furthermore, this MSE is mainly constant, having in mind that this algorithm did manage to represent the original shape of signal, with aforementioned flaw about latency.

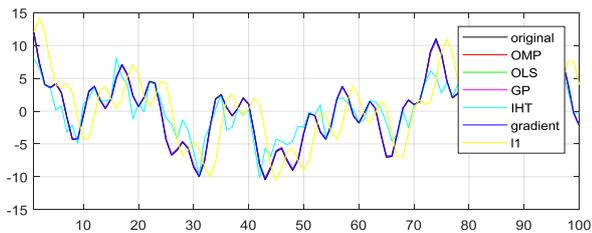

Figure 3: Time domain of original and reconstructed signals

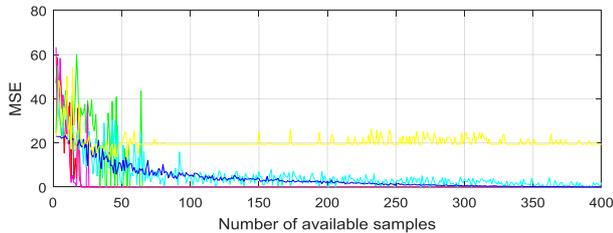

Figure 4: MSE dependency on available samples

## V. CONCLUSION

In this paper, we executed a comparison between six sparse recovery algorithms, mainly from Greedy and Convex optimization family and one from Thresholding family. The experimental results show that Greedy and Convex algorithms have edge over others, with lower reconstruction time and MSE. They all had similar performances which outshadowed IHT and l1-eq. Despite all those advantages, Greedy algorithms have weaknesses due to the fact that the exact number of signal components must be known before reconstruction occurs.